# Quantum Machine Learning: Quantum Kernel Methods

Sanjeev Naguleswaran, QSPectral Systems

## Introduction

Quantum algorithms based on quantum kernel methods have been investigated previously [1]. A quantum advantage is derived from the fact that it is possible to construct a family of datasets for which, only quantum processing can recognise the intrinsic labelling patterns, while for classical computers the dataset looks like noise. This is due to the algorithm leveraging inherent efficiencies in the computation of logarithms in a cyclic group.  The discrete log problem.is a well-known advantage of quantum vs classical computation: where it is possible to generate all the members of the group using a single mathematical operation.

Kernel methods are a powerful and popular technique in classical Machine Learning. The use of a quantum feature space that can only be calculated efficiently on a quantum computer potentially allows for deriving a quantum advantage [1]. In this paper, we first describe the application of such a kernel method to a Quantum version of the classical Support Vector Machine (SVM) algorithm to identify conditions under which, a quantum advantage is realised. A data-dependent projected quantum kernel was shown to provide a significant advantage over classical kernels [2].

In this paper, we use data from the Venture Capital (VC) investment process to determine the validity of using quantum kernel methods. Thus far, other than for a few notable exceptions, deciding on companies that are likely to succeed is undertaken through labour-intensive screening processes and the intuition of the investor [3][4].

We present the results of investigations and ideas pertaining to extending the use of quantum kernels as a feature extraction layer in a Convolutional Neural Network (CNN), which is a widely used architecture in deep-learning applications. In particular, we discuss,

- The investigation and development of quantum kernel functions that leverage quantum-enhanced feature spaces for favourable representations of real-world data; and
- Investigation of properties and structure of real-world data that could be leveraged to provide a quantum advantage.

## Method

Machine Learning (ML) is a component Artificial Intelligence (AI) that focuses on learning from data. It primarily addresses the interface between learning, complexity, and computation. Research in this area spans many theoretical topics as well as practical applications. ML concerns the extraction of insights from data, developing predictions to permit data-driven decision-making. Algorithms that constitute ML fall into two broad categories: Supervised and Unsupervised learning. Supervised learning is applicable when training data is available and typically most classification and regression algorithms fall within this category. Decision Trees, Neural Networks Support Vector Machines (SVM) and Bayesian networks can also be used under supervised learning. In the case of unsupervised learning clustering algorithms, such as k-means, clustering can be used to derive insights into the nature of the data. In practice, other categories such as semi-supervised learning are encountered where the boundary between algorithm types is blurred, resulting in hybrid fit-for-purpose algorithms using any of the available approaches in an appropriate manner. Machine Learning theory is also closely connected to statistical inference where a basis function is applied to combine the features of a problem to infer the response for the subset of data used for training. Further details on Machine Learning can be found in [5], [6].

The start-up picking (or identification) problem can be considered as classifying or clustering the start-ups into predefined categories. Machine Learning provides several methods with varying degrees of sophistication and complexity to achieve this classification and/or clustering objectives.

Decision trees are decision support tools based on a graph structure where the nodes are tests on an attribute (feature) of a problem. The attributes particular to investment data could include the chance of an event or measurement, measurement thresholds and the industry domain. Decision tree learning can be used as a predictive model to map features or observations to a target value. In particular, a classification tree is defined as a decision tree where the target variable has a finite set of values (or classes). In contrast in a Regression Tree, the target variable is continuous and real. Therefore, the resulting decision trees can be used to predict the value of a target variable based on several input variables [7].

A further enhancement to ML is the use of Ensemble methods that take several simple models and combine them in a way to yield a final, overall picture. In our benchmarking, we used an ensemble technique based on Random Forest models that iteratively averaged multiple, deep decision trees, which are trained on different parts of the collected data, to reduce the variance. Each iteration creates a simple decision tree on randomly selected subsets of input variables and input data. The result is formed by the aggregation of all such trees [8].

Support Vector Machines (SVMs) are another type of supervised learning algorithm that can be used for classification or regression tasks. They are based on the idea of finding a hyperplane in high-dimensional space that maximally separates different classes. The SVM algorithm separates the classes (such as successful start-up vs unsuccessful) by drawing a hyperplane between the data points [9]. In cases where the data cannot be linearly separable, because the algorithm is based on kernel methods, a mathematical concept known as the "kernel-trick" is of particular interest in extending the method to leverage Quantum Computing.  The kernel trick works by applying a non-linear transformation to the data before it is fed into the algorithm. This transformation projects the data into a higher-dimensional space, where it may become linearly separable. The algorithm can then be applied to the transformed data, and the resulting model can be used to make predictions on the original data [10].

The kernel trick allows SVMs to handle non-linear classification tasks by implicitly mapping the data into a higher-dimensional space using a kernel function. This allows the algorithm to find a non-linear decision boundary that can accurately separate the classes in the data. However, when the feature space is large the method encounters limitations due to the computational complexity of estimating the kernel functions. The use of a quantum feature space that can only be calculated efficiently on a quantum computer potentially allows for deriving a quantum advantage. Previous work on quantum kernel methods used in ML is described in [1]. Further, an important approach in the search for practical quantum advantage in machine learning is to identify quantum kernels for learning problems that have underlying structure in the data. Examples of learning problems for data with group structure have been identified and a class of kernels related to covariant quantum measurements were constructed [10]. In addition, a data-dependent projected quantum kernel was shown to provide significant advantage over classical kernels [2].

A barrier to leveraging ML methods is the scarcity of data and this would affect the smaller VC firms who would lack internal data that the big firms (such as Google Ventures or Sequoia Capital) can access through their internal processes and their associated business applications. Paradoxically, these small firms are the ones that would benefit more from AI-driven decision support as they would normally lack the resources to employ a team of analysts.

It was also imperative to investigate and develop Machine Learning methods to operate on the real-world datasets pertaining to the start-up problem. To train the machine learning model, we used investment data about start-up companies obtained from Crunchbase [11].

The original data contained 49437 companies (rows) and 39 features (columns). However, as in this work only successful (exited) vs unsuccessful (closed) companies are considered (i.e., the operational companies were not considered in this work) this was reduced to 5497 companies. Due to determining correlations between features and considerable effort in cleaning the dataset, the dataset was also consolidated into 17 features through feature engineering. The dataset contained company information such as, name of the company, URL, market, country, state, region, city, founded date, first funding date, last funding date. It also had data on different investment types such as seed, venture equity crowdfunding, undisclosed funding, convertible note, debt financing, angel, grant, private equity, post IPO equity, post IPO debt, secondary market, product crowdfunding, round A-H series funding.

It should be noted this dataset is not perfectly structured and complete like the ideal datasets that algorithm performance is usually evaluated against. In the first instance, we benchmarked performance using the Decision Tree and Random Forest algorithms.

After the performance parity of the SVM was established a quantum kernel method was developed using the Pennylane Quantum Application Programming Interface (API) [12] - this kernel was then substituted for the classical kernel.

## Results and Implications

The ML performance on this dataset was first benchmarked by developing and optimising a Decision Tree and a Random Forest algorithm. The performance of the algorithms can be evaluated by means of the accuracy report where a probability as close as possible to one (100%) is desirable (a probability of 0 is means there is no predictive capability and 0.5 would be analogous to flipping a coin). The following accuracy reports shown in Figures 1 and 2 were obtained:

```
                precision    recall  f1-score   support

     Class 0       0.60      0.58      0.59       432
     Class 1       0.73      0.75      0.74       668

    accuracy                           0.68      1100
   macro avg       0.67      0.67      0.67      1100
weighted avg       0.68      0.68      0.68      1100
```

*Figure 1 Classification report for Decision Tree algorithm*

```
                precision    recall  f1-score   support

     Class 0       0.61      0.54      0.57       432
     Class 1       0.72      0.78      0.75       668

    accuracy                           0.68      1100
   macro avg       0.67      0.66      0.66      1100
weighted avg       0.68      0.68      0.68      1100
```

*Figure 2 Accuracy report for Random Forest*

Where the classification accuracy report is a summary of the performance of a classifier on a test set. It includes the following metrics:

1. Accuracy: This is the ratio of the number of correct predictions made by the classifier to the total number of predictions. It is defined as:

Accuracy = (True Positives + True Negatives) / Total

2. Precision: This is the ratio of true positive predictions to the total number of positive predictions made by the classifier. It is defined as:

    Precision = True Positives / (True Positives + False Positives)
3. Recall: This is the ratio of true positive predictions to the total number of actual positive instances in the test set. It is defined as:

    Recall = True Positives / (True Positives + False Negatives)

4. F1 Score: This is the harmonic mean of precision and recall. It is a balanced measure of the classifier's performance, considering both the precision and recall. It is defined as:

    F1 Score = 2 * (Precision * Recall) / (Precision + Recall)

5. Support: This is the number of samples of each class in the test set.

The classification accuracy report is an important tool for evaluating the performance of a classifier and for comparing different classifiers. A classifier with high accuracy, precision, and recall is generally considered to be performing well. However, it is important to note that the choice of evaluation metrics can depend on the specific goals and requirements of the task being attempted.

In the above report where class 1 represents the companies that had a successful exit and class 0 represents the companies that have not exited (or closed), it should be noted that 75% of the successful companies have been correctly identified and 72% of the companies identified as successful are correct. The overall accuracy of the prediction model is 68%.

A comparison of the performance of an SVM algorithm to the above benchmark resulted in a very similar outcome. The results obtained are shown in Figures 3 and 4.

|           | precision | recall | f1-score | support |
|-----------|-----------|--------|----------|---------|
| Class 0   | 0.57      | 0.66   | 0.62     | 432     |
| Class 1   | 0.76      | 0.68   | 0.72     | 668     |
|           |           |        |          |         |
| accuracy  |           |        | 0.67     | 1100    |
| macro avg | 0.67      | 0.67   | 0.67     | 1100    |
| weighted avg | 0.69   | 0.67   | 0.68     | 1100    |

*Figure 3 Accuracy report for Support Vector Machine*

Where the accuracy report is comparable to the Random Forest algorithm. In this case, a confusion matrix, which is a table that shows the number of true positive, false positive, false negative, and true negative predictions made by the classifier for each class is also shown in Figure 4.

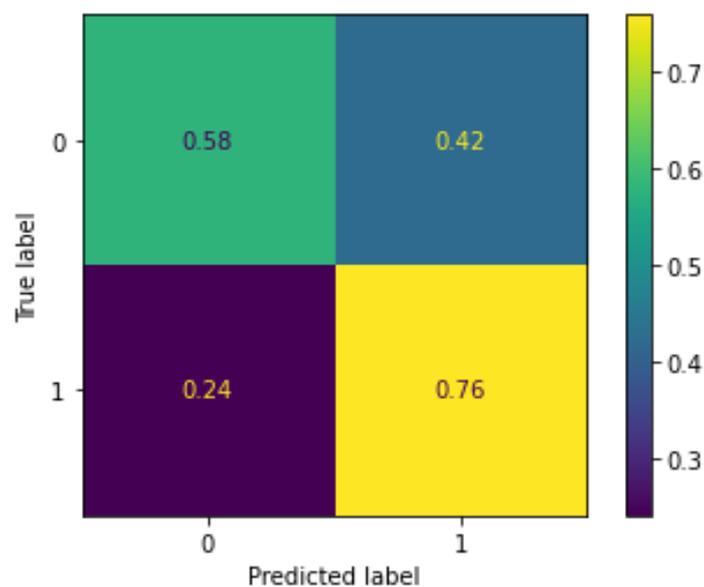

*Figure 4 Confusion matrix for SVM showing the correct and misrepresentation of classes)*

A quantum kernel was then developed and substituted for the classical kernel in the algorithm with the following outcome.

|  | Precision | Recall | F1-Score | Support |
|---|---|---|---|---|
| Class 0 | 0.58 | 0.52 | 0.55 | 432 |
| Class 1 | 0.72 | 0.78 | 0.75 | 668 |
|  |  |  |  |  |
| accuracy |  |  | 0.66 | 1100 |
| macro avg. | 0.64 | 0.64 | 0.64 | 1100 |
| weighted avg. | 0.68 | 0.66 | 0.66 | 1100 |

*Figure 5 Accuracy report for Quantum SVM*

It can be seen that that performance is comparable to classical SVM. This is encouraging as even without imposing a group structure on the data there is no degradation in performance. The outcome is also represented in the confusion matrix shown below in Figure 6.

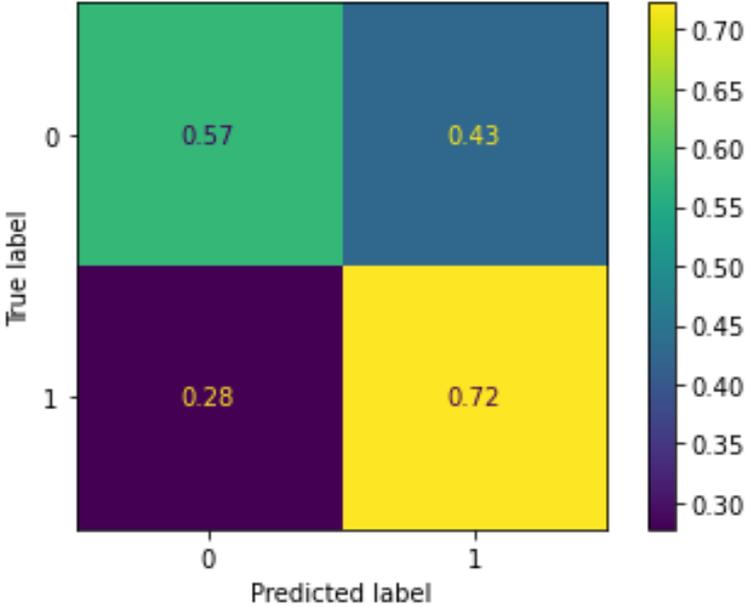

*Figure 6 Confusion matrix for Quantum SVM*

Motivated by this result, A further experiment where a quantum convolutional layer was added to a classical Neural Network as a pre-processing layer was conducted. The following comparisons in accuracy with and without the quantum kernel are shown in Figure 7:

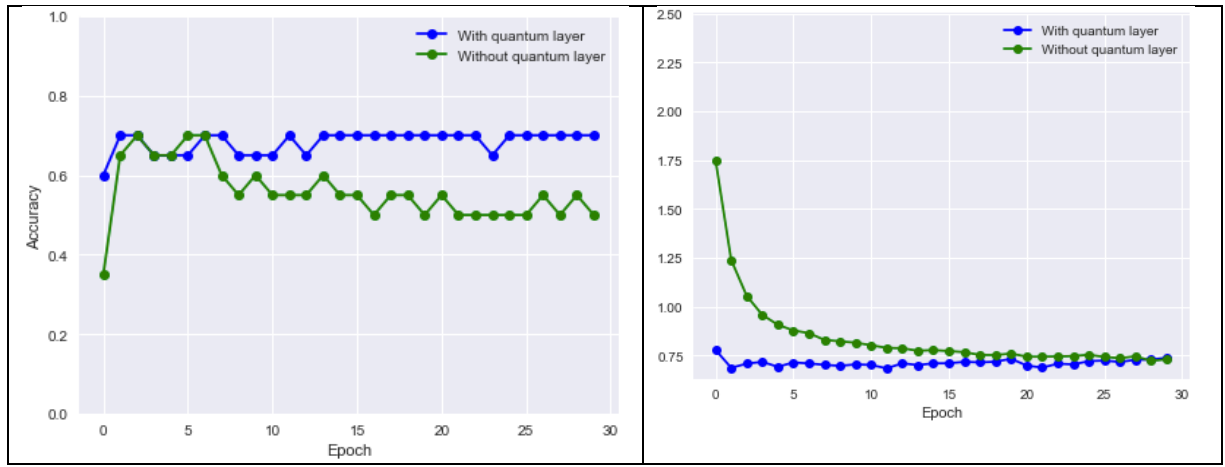

*Figure 7 Comparison of Accuracy and Loss after including quantum pre-processing*

It was shown that when the quantum kernel method was used in a hybrid quantum-classical Convolutional Neural Network there was a significant quantum advantage.

## Conclusion

In this work, we have shown the feasibility of using quantum kernel-based Machine Learning on a Real-World non-standard dataset. performance to classical ML was established in the fundamental case of applying an SVM algorithm without restructuring data. Additionally, when a quantum convolutional kernel was used as a pre-processing layer in a hybrid quantum-classical Neural Network an advantage over a purely classical network was observed with the hybrid Neural Network.

Previously, we presented a paper showing that it is possible to employ a simulation-based approach to venture portfolio optimisation [13]. However, the missing aspect of the problem was the identification of start-ups to include in the simulated portfolio with adequate mathematical rigour. In this paper, we present a machine-learning-based solution enhanced by quantum methods to address this outstanding issue. The combination of the start-up selection and portfolio optimisation methods would result in a level of sophistication and rigour employed in constructing portfolios of more traditional asset classes and therefore, would satisfy the risk profiles of a more diverse investment pool.

# References

1. Havlíček, V., Córcoles, A. D., Temme, K., Harrow, A. W., Kandala, A., Chow, J. M., & Gambetta, J. M. (2019) Supervised learning with quantum-enhanced feature spaces. *Nature*, 567, 209.
2. Huang, H. Y., Broughton, M., Mohseni, M., et al. (2021) Power of data in quantum machine learning. *Nature Communications*, 12(1), 2631.
3. Zider, B, "How Venture Capital Works", *Harvard Business Review*, v.76, no.6 ,1998
4. Bhakdi, J. (2013) Quantitative VC: A new way to growth. *The Journal of Private Equity*, 17(1), 14–28.
5. Bishop, C. M. (2006) *Pattern Recognition and Machine Learning*. Springer.
6. Hastie, T., Tibshirani, R., & Friedman, J. H. (2001) *The Elements of Statistical Learning*. Springer.
7. Breiman, L., Friedman, J. H., Olshen, R. A., & Stone, C. J. (1984) *Classification and regression trees*. Monterey, CA: Wadsworth & Brooks/Cole Advanced Books & Software.
8. Breiman, L. (1996) Bagging predictors. *Machine Learning*, 24, 123–140.
9. Vapnik, T. E., & Poggio, M. (2001). *Support Vector Machines: Theory and Applications*. 2049, 249-257.
10. Glick, J. R., Gujarati, T. P., Córcoles, A. D., Kim, Y., Kandala, A., Gambetta, J. M., & Temme, K. (2021) Covariant quantum kernels for data with group structure. *arXiv:2105.03406*.
11. Crunchbase. Available at: https://www.kaggle.com/arindam235/startup-investments-crunchbase, accessed December 2022.
12. Pennylane. Available at: https://pennylane.ai/, accessed December 2022.
13. Naguleswaran, S. (2021) Quantum Computing for Simulating Start-ups: Investor Portfolio Optimisation. *ACERE Conference*, February.